\documentclass[12pt, letter paper]{article}
\usepackage[section]{placeins}
\usepackage{graphicx}
\graphicspath{{images/}}
\title{Proposed experimental test of Randall-Sundrum Models}
\author{Behnam Pourhassan$^1$, Anha Bhat$^2$, Hrishikesh Patel$^3$,\\ Mir Faizal$^{3,4}$,  Nicholas Mantella$^3$ \\\\
$^1$School of Physics, Damghan \\ Universitty, Damghan, 3671641167, Iran\\
$^2$Department of Electronic Materials Engineering,\\
Research School of Physics and Engineering,\\Australian National University,
Canberra,\\ ACT 0200, Australia\\
$^3$Irving K. Barber School of Arts and Sciences,\\  University of British
Columbia - Okanagan,\\   Kelowna, British Columbia V1V 1V7, Canada \\
$^4$Department of Physics and Astronomy, \\
University of Lethbridge, Lethbridge,\\ AB T1K 3M4, Canada   \\
}
\date{}
\begin{document}
\maketitle

\begin{abstract}
 The Randall-Sundrum models are expected to modify the short distance behavior of  general relativity. In this paper, we will propose an  experimental test for this short distance  modification due to Randall-Sundrum models. This will be done by analyzing motion of  a particle which is moving  in  spherical gravitational field with a  drag force.  The  position at which the particle stops  will be different in general relativity and Randall-Sundrum model. This difference in the distance moved by the particle before stopping  can be measured using a Nanoelectromechanical setup. Thus, it is possible to experimentally test Randall-Sundrum models using currently available technology. 

\end{abstract}

\section{Introduction}

The smallest scale at which gravity has been tested is  $0.4mm$ \cite{mz4}-\cite{mz5}.  It has been observed that general relativity (and Newtonian gravity as its short distance approximation) holds till this scale. However, as general relativity is only an effective field theory approximation to string theory, and if string theory is the actual theory describing nature, then it is expected that general relativity will be corrected at short distances. This short distance correction to general relativity will cause short distance corrections to the Newtonian gravity. Such a modified Newtonian gravity can be detected in near future experiments. So, it is important to consider various corrections to the Newtonian gravity at short distances from string theory. 

It is known that string theory predicts the existence of extra dimensions, and this has motivated the construction of brane world models \cite{b1}-\cite{b2}. In these brane world models, it is predicted that our universe is a brane in a higher dimensional bulk. The existence of these extra dimensions can lower the effective Planck scale by several orders of magnitude, and this can have important low energy consequences \cite{b5}-\cite{b6}. The Randall-Sundrum (RS) models are brane world models, with warped geometry, such that the matter fields are localized on a four dimensional brane, and the gravity can move in the bulk \cite{r1}-\cite{r2}.

It has been possible to relate the Randall-Sundrum models to a particular compactification of Type IIB string theory \cite{iib1}-\cite{iib2}. In this approach,   Randall-Sundrum models are obtained from a combination of stack of negative tension branes stuck at an orbifold fixed point. In fact, the warping
scale is ${(4 \pi g N)^{1/4}}/{ M_{st}}, $ where $g$ is the string coupling, $M_{st}$ is the string scale, $N$ is the  the number of branes in type IIB string theory, which have been stacked to form the RS brane \cite{iib1}-\cite{iib2}. Now, it can be observed that by considering a large number
of such branes in type IIB string theory, the warping scale for  Randall-Sundrum model can be lowered, and this can modify the low energy behavior of such
systems  \cite{ib12}-\cite{ib14}. 
It is possible to relate  the world-volume fields of probe D7 branes to  bulk fields in Randall-Sundrum  models \cite{probed}.  This is done using the Klebanov-Witten background of type IIB string theory. In their near-horizon geometry, these  constructions are described by $AdS_5 \times T^{1,1}$. The $\mathcal{N} =1$ supersymmetry of the dual theory is produced from   $T^{1,1}$, which  is  a five dimensional  compact internal manifold.

The Randall-Sundrum models have  been used to study interactions between   dark matter particles and standard model  particles \cite{darkmatter}. These dark matter  particles interact through the exchange of spin two Kaluza-Klein  gravitons   with the standard model particles.  
It has also been predicted that such Randall-Sundrum  models can modify the Hawking radiation emitted by black holes, which can be detected from gravitational wave observations \cite{gw}. It has also been predicted that the   Newtonian potential would get  corrected in the  Randall-Sundrum models at short distances \cite{new1}-\cite{new2}. This short distance modification of Newtonian potential can be detected and this in turn can be used to test the Randall-Sundrum models. So, in this paper, we will propose an experimental setup for the detection of these corrections from Randall-Sundrum models. 

\section{Potential in Randall-Sundrum  Models}
In this section, we will discuss the corrections to Newtonian potential from Randall-Sundrum models. 
The action in Randall-Sundrum  models can be written (using the five dimensional scalar curvature $R_5$, five dimensional metric $g_5$, five dimensional Planck mass $m_5$, five dimensional cosmological constant $\Lambda_5$) as
\cite{new1}-\cite{new2},
\begin{equation}
S = \int d^4 x dy \sqrt{|g_5|}(2 m_5^2 R_5 - \Lambda_5) + \int d^4 x \sqrt{|g_4|} \mathcal{V},
\end{equation}
where the four dimensional metric $g_4$ is the metric on the hyper-surface $y=0$, and $\mathcal{V}$ is  the brane tension.
The solution to this brane world model can now be written as
\begin{equation}
d s^2 _5 = e^{|y|/l} ds^\mu ds^\nu g_{\mu\nu}- dy^2,
\end{equation}
where $g_{\mu\nu} = \eta_{\mu\nu}$ is the four dimensional metric, and $l$ is
is the  radius of the  bulk AdS spacetime. Here, the brane tension, and $\Lambda_5$ are given by
\begin{eqnarray}
\mathcal{V} = \frac{24 m_5^3}{l}, && \Lambda_5 =  - \frac{24 m_5^3}{l}
\end{eqnarray}
Thus, the four dimensional cosmological constant vanishes. The brane world model can be analyzed using this action, and various consequences of such a model  have been derived \cite{new1}-\cite{new2}.
The short distance modification to Newtonian potential is obtained by analyzing the modification of effective gravitational constant with length scale in Randall-Sundrum models.

So, the Newtonian potential gets corrected in the brane world models at short distances, and such corrections for Randall-Sundrum models can be written as \cite{new1}-\cite{new2},
\begin{equation}\label{1}
\Phi(r)=-\frac{GM}{r} \left( 1+\frac{k}{r} \right),
\end{equation}
where $k$ determines the length scale at which corrections due to Randall-Sundrum models become significant. These corrections become relevant in the sub-millimeter range \cite{new1}-\cite{new2}. We  know that the Newtonian gravity has been probed and verified to the scale of $0.4mm$ \cite{mz4}-\cite{mz5}. A bound on corrections to such brane world models has also  been obtained   \cite{new1}-\cite{new2}, and it is about $  0.169mm$. Thus, if we design an experiment capable of probing gravity at $0.1mm$ or below, we might be able to create a stringent bound on the size of extra-dimensions predicted by Randall-Sundrum models and also test whether Newtonian gravity is modified below the sub-millimeter scale.  
It may be noted that we can analyze the modification of dynamics classically, as quantum effects can be neglected at this scale.

\section{Experimental Setup}

In this experiment, we first construct a sphere with a narrow hollow tube passing through one of it's diameters. The sphere is much larger than the hollow tube, and as such the effect of the hollow tube on the gravitational field of the sphere can be neglected. Now we let a particle free fall in this sphere, and it falls under the gravitational field of this sphere. We analyze the dynamics of this particle, in both Newtonian gravity, and in the Randall-Sundrum corrected gravity. It is observed that the dynamics of the free falling particle changes, and this can possibly be detected. This changed dynamics of the particle can be used to determine if such corrections to the Newtonian gravity occur at short distances. We will demonstrate that for the Newtonian potential, the particle would perform harmonic oscillation around  the center of the sphere.  So, if a particle is released from one end of the sphere than it will oscillate with an amplitude equal to the radius of the sphere. However, this situation gets corrected in brane world models due to a shift in the equilibrium point. However, because the length scale in use is very small, no significant deviation might be observed by direct measurement of this motion. So, we put a viscous medium in the diametrical canal which can be used to damp the motion leading the particle to eventually stop it.  Now as the Newtonian law is expected to get corrected, a particle following the Newtonian law will stop at a different place as compared to a particle following the Randall-Sundrum  modified Newtonian law. We intend to measure this deviation in distance to test  the law this particle actually follows in nature.

Hence, we present a setup in which the radius of the sphere is of the order of the Randall-Sundrum length scale $(k \approx{} 0.1mm)$ because that is the maximum order at which the corrections are predicted to occur. The particle should be as small as possible to prevent the coupled motion of the system about their centre of mass. Hence, we choose the particle size to be hundredth of the size of the sphere to minimize this coupling.

\begin{figure}[h!]
\begin{center}
$%
\begin{array}{cccc}
\includegraphics[width=90 mm]{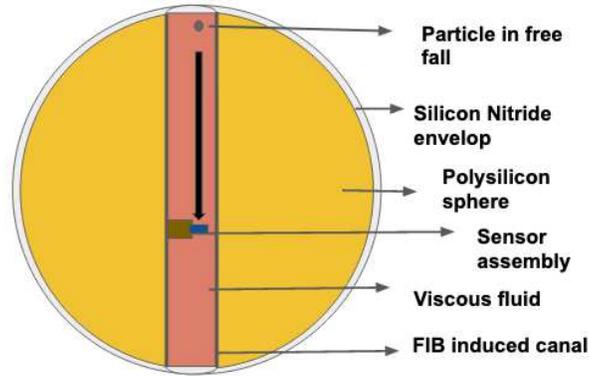}
\end{array}%
$%
\end{center}
\caption{Simplistic model of the sphere}
\label{fig3-1}
\end{figure}

The sphere with these dimensions of mass $0.078 mg$ can be made out of high density poly-crystalline silicon. The sphere is enveloped within silicon nitride except the circular opening on top of it which is about $2 \mu m$ wide. The particle of size $1 \mu m$ is released in free fall from the top of the canal which has a viscous fluid in it. A focused ion beam (FIB) is used to mill a cylindrical canal of $ 200 \mu m$ length across the sphere. The FIB assembly comprises of the electrodes which operate in the voltage range of $5-50 KeV$ and are connected to the tungsten tip located in the reservoir of the liquid metal (gallium) ion reservoir. The whole assembly is at a distance of $200 \mu m$ from the polysilicon sphere. The gallium ions are pulled up in a sharp tungsten cone under vacuum having the radius of $5-10 nm$ and are bombarded towards the circular aperture. The repetitive influx of the gallium ions etched away the polysilicon through the entire diameter of the sphere. Some of the polysilicon melts within the canal which also contains the gallium residue which can be flushed out with $XeF_2$ etching solution. As a result, a canal of the order of $200 \mu m$ running across the sphere from top to the bottom is obtained.

\begin{figure}[h!]
\begin{center}
$%
\begin{array}{cccc}
\includegraphics[width=90 mm]{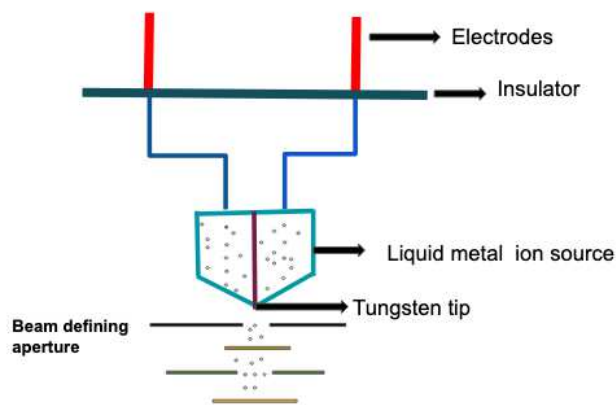}
\end{array}%
$%
\end{center}
\caption{FIB setup}
\label{fig3-2}
\end{figure}

\begin{figure}[h!]
\begin{center}
$%
\begin{array}{cccc}
\includegraphics[width=90 mm]{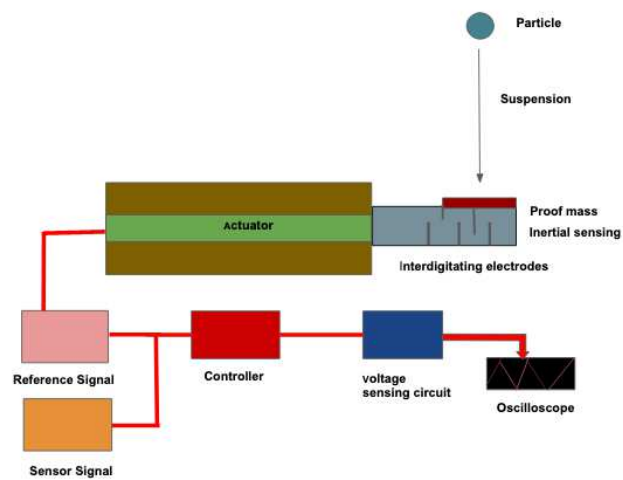}
\end{array}%
$%
\end{center}
\caption{NEMS assembly}
\label{fig3-3}
\end{figure}

Once the tunnel is etched, a nanoelectromechanical sensor (NEMS) assembly is placed at a distance of $ 150 \mu m$ from the top. The NEMS comprises of a proof mass-based platform supported on interdigitating electrodes. The electrodes act as charge storing capacitors which maintains a uniform electrostatic field, except when it experiences changes in the field based on the inertial differences on the platform above it. The changes are picked up by the actuator which sets the feedback loop into action and the responses are picked up based on the inertial sensing. The inertial sensing is picked up when the particle strikes the proof mass surface, and  in other case it does not show any difference. The apprehension of the parasitic capacitance is calibrated before the experiment is conducted. The particle follows two main courses of action. It either hits the sensor platform moving pass the centre or it gets stuck in the middle. In either of the cases, the platform initiates the reference signal or generates the sensor signal which is detected by the effector loop. The assembly which is initially set at $150 \mu m$ from the top can be further adjusted to a distance where we theoretically predict the particle to stop.

\section{Motion of the Particle}

\begin{figure}[h!]
\begin{center}
$%
\begin{array}{cccc}
\includegraphics[width=80 mm]{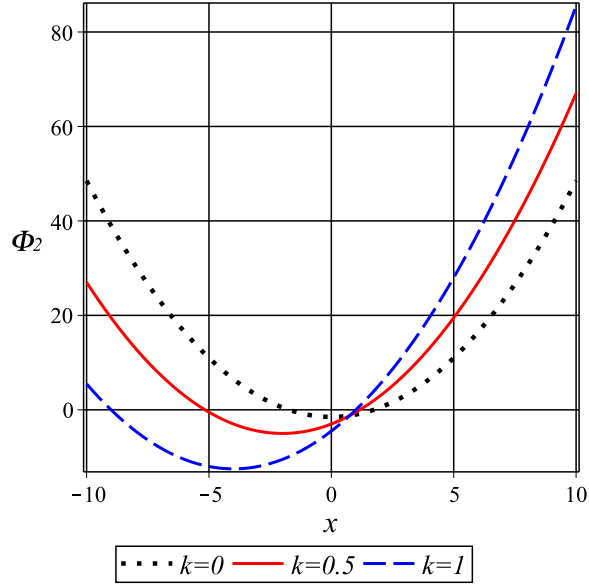}
\end{array}%
$%
\end{center}
\caption{Typical behavior of modified potential for $R=M=G=1$.}
\label{fig1}
\end{figure}

In this section, we will calculate  the  distance moved by the particle in Randall-Sundrum models and general relativity. 
Now we can write  Newtonian potential for a particle, which is oscillating in a sphere of mass $M$ as
\begin{equation}\label{2}
\Phi_1 = -\frac{GM}{2R^3}(3R^2-x^2).
\end{equation}
The Randall-Sundrum correction to this  potential, for such a particle can be written as
\begin{equation}\label{3}
\Phi_2 = -\frac{3GM}{R}\left(\frac{1}{2}+\frac{k}{R}-\frac{4k x}{3R^2}-\frac{x^2}{6R^2}\right).
\end{equation}
We can use this corrected potential to analyze the dynamics of this system.
It is clear that the minimum of potential occurs at $x=-4k$, which can be  interpreted as stable equilibrium point for the particle. Hence the particle  oscillates around the point $x=-4k$. In Fig. \ref{fig1}, we can observe the variation of potential $\Phi_2$, with the correction parameter $k$. In the case of $k=0$, particle can oscillate around the origin. Effect of $k$ shifting the equilibrium point is illustrated in the Fig. \ref{fig1}. Dotted black point with minimum at ($x=0$) represent $\Phi_1$. Value of the modified potential at the $x=-4k$ is $\Phi_2(x=-4k)=-\frac{3}{2}-3k-8k^{2}$ (with $R=G=M=1$). Therefore,   at  $k=-\frac{x}{4}$, the force acting on the particle vanishes. As it is an equilibrium point with negative $x$, the value of $k$ is positive.

Now, we can write the Lagrangian for the motion of the particle in corrected Newtonian law as,
\begin{equation}\label{4}
L(x,\dot{x})=\frac{1}{2}m\dot{x}^2-\Phi_2(x).
\end{equation}
Hence, we can write the solution for the Randall-Sundrum models as
\begin{equation}\label{9}
x = \frac{1}{\omega}\sqrt{16k^2\omega^2+2c}\sin(\omega t) - 4k,
\end{equation}
where $\omega =  {GM}/{R^3}$, and we set the phase angle as $\theta_{0}=0$.
Similarly, the velocity ($v$) of the particle is   $v = \sqrt{16k^2\omega^2+2c}\cos(\omega t)$.
At $k=0$, the modified potential reduces to the Newtonian potential, and the equation for the motion of the particle in the modified gravitational potential reduces to the usual case. Here, $c$ is an arbitrary parameter, which controls the amplitude of the oscillator, so we can choose it in such a way that the amplitude of oscillation is equal to the radius of the sphere. In order to create an appreciable deviation in the amplitude of the particle, a relatively small value of $c$ has to  be taken, and thus a smaller value of $R$ could help us detect the deviation.\par

The derivation of the parameters for the motion of the particle in the case of damping has been provided in the appendix. Provided below is the approximate time and length the particle will cover before coming to a complete stop due to the damping provided by a fluid of viscosity $b$ that is filled in the tunnel.\par

For the small $b$ ($b<<\omega$) the time to stop the motion $(t_{t})$ can be approximated as,

\begin{equation}\label{t}
t_{t}\approx{\frac {1}{\sqrt {{\omega}^{2}-{b}^{2}}}\tan^{-1} \left( {\frac {\sqrt
{{\omega}^{2}-{b}^{2}}}{b}} \right) },
\end{equation}
$(t_{t})$ is clearly independent of correction parameter $k$. The Distance travelled before stopping depends upon the correction parameter as,
\begin{equation}\label{x}
x\approx \frac {{\xi}\sqrt {{\omega}^{2}-{b}^{2}}}{\omega^{2}} \left(1 -{\frac {b}{\sqrt {{\omega}^{2}-{b}^{2}}}
\tan^{-1} \left( {\frac {\sqrt {{\omega}^{2}-{b}^{2}}}{b}} \right) } \right)-4\,k, 
\end{equation}
where
$
\xi =\sqrt{16k^2\omega^2+2c}.
$
In Eq. \ref{x}, we observe that the total distance travelled before the particle stops in terms of (a) correction parameter and (b) viscous drag coefficient. In both cases, we can see critical value for the drag coefficient, where $x=0$ is given by,
\begin{equation}\label{x=0}
b_{c}\approx {\frac {\omega\, \left( \pi {\xi} +
\sqrt { \xi^2 \left( {\pi }^{2}-8 \right) +32k\omega {\xi}} \right)
}{2 {\xi}}}.
\end{equation}

Theoretically it seems that the proposed setup is viable, however, it is important to stop and consider some practical factors that might affect the system.  Starting with the viscosity of the medium filled in the tube. If we take $c=0.5$ and $\omega=1$, one can obtain $b_{c}\approx2.2$ for $k=0$. As $k$ is very small, the value of $b_c$ does not change considerably with the modified law. However, it is observed that $b_c$ and $\omega$ have an almost linear relationship. It is also observed that at greater values of $\omega$ the deviation in the motion described by the classical law and the modified law also increases. However, the homogeneity of the medium also needs to be controlled to obtain accurate measurements. Thus, it is important to properly adjust the  viscosity of the medium to be filled in the diametrical canal of the sphere. \par

The value of $b_c$ can be chosen such that the motion described by classical and modified law are either critically damped or over-damped. Here, $b_c$ has been calculated such that the particle performing the motion described by the modified law stops due to the damping force at $x=0$, which is the centre of the sphere. A particle obeying the Newtonian  law will also stop due to over-damping, however, it will travel an additional distance $A$ before stopping, this difference in the position can be measured. This can be used to verify if the predicted power law modifications to Newtonian gravity actually occur. \par 

\section{Errors}

It is important to note that due to the robustness and the simplicity of the setup, there are very few modes in which errors and uncertainties could affect the system. As we use a nanoelectromechanical system (NEMS) to detect where the particle hits the interdigitating electrodes, we should be able to measure deviations up to a scale of at least a few micrometers confidently. This is comparable and in many cases better than the setups proposed in previous work in this field \cite{mz4}-\cite{mz5}. Many conventional setups are based on the Cavendish experiment which has a considerable scope of errors. The proposed setup is advantageous over the conventional methods due to its simplicity and its ability to make accurate measurements. \par 

However, the experiment is not completely devoid of noise and thus there are still some sources of errors. The errors in the system due to electromagnetic forces generated due to the internal components of the setup can be significantly reduced by using dampeners, which can reduce any residual electric field created by the NEMS. Furthermore, material of the oscillating particle can also be chosen such that the small electric field created by the NEMS does not induce charge on the particle. This can significantly limit the strength of any electromagnetic interactions. The quantification of these forces depends on the type of fluid we fill into the canal. The choice of fluid depends on its viscosity, which also affects the homogeneity of the setup. Thus, quantification of the forces is not possible until a fluid is chosen.  
However, with the use of dampeners, proper choice of material for the particle, and with some of the methods suggested in \cite{mz5}, we can substantially reduce the electromagnetic effects in order to make precise measurements at the scale of few micrometers. Furthermore, as the tunnel is proposed to be narrow, we anticipate Casimir forces due to residual electromagnetic forces on the particle due to the sphere. It has been observed that Casimir forces of the order $10^{-13} N$ occur at a distance scale of $10^{-7} m$ \cite{casimir}, and as the distance scale between the particle and the sphere is greater than that, we expect even smaller Casimir effects. As these accelerations scales are below the range we anticipate to probe, Casimir effects can be neglected. \par

Furthermore, in order to avoid significant noises and a significant impact of earth's gravity on the system, the experiment has to be done in the outer space, most probably on a satellite. As the system will still be under the influence of earth's gravity, we expect tidal forces to influence the system. However, if we keep the system at a distance from earth that is equal to the distance between the earth and the moon, the tidal accelerations are of the order of $10^{-15} m/s^2$, which is much smaller than the accelerations we are measuring, and hence they can be neglected. \par

Now, it is important to note that the proposed setup is ideal for power law corrections to Newtonian gravity. As Randall Sundrum models predict such modifications at a scale below $0.169 mm$ \cite{new2}, and because conventional methods have only been able to test gravity until $0.4mm$ \cite{mz4}-\cite{mz5}, the conventional methods could not be used to test the validity of Randall-Sundrum models. However,  our setup can be used to    qualitatively verify/falsify the Newtonian law below this scale at which the Newtonian potential is corrected  by the Randall-Sundrum models.  Furthermore, as  the errors in our system are much smaller than the scale at which this potential needs to be tested, it is possible to test the experimental validity of Randall-Sundrum models using our experimental  setup. \par

\section{Conclusion}

It is known that the  Randall-Sundrum models are expected to modify the short distance behavior of  general relativity. This modification is expected to occur at $0.169 mm$, and gravity has been presently tested only till $0.4mm$. So, it is possible to test  Randall-Sundrum models by analyzing  gravitational potential at a slightly smaller scales. In this paper, we propose to  analyze the  motion of  a particle  in  spherical gravitational field with a  drag force.  As the gravitational potential is corrected in Randall-Sundrum model, the particle will travel  different distances before stopping in  Randall-Sundrum model and general relativity. Measuring this distance involves the use of a Nanoelectromechanical setup. We make use of such a setup as the errors in NEMS are much less than the scale at which the gravitational potential has to be measured in this experiment. Thus, this present experimental setup can be effectively used to test the Randall-Sundrum models and other power-law modifications to gravity in the sub-millimeter range.    \par

It may be noted that this setup, and certain modifications of this setup can be used to test several other forms of short distance modification of Newtonian potential, which occur in various different models. 
There are several different forms of modification to Newtonian potential which can be obtained using different brane world models.
The graviton propagator in the Randall-Sundrum model with the Gauss-Bonnet interaction has been studied, and it has been observed that
the ghost-free condition is sufficient for obtaining the Newtonian gravity. Randall-Sundrum geometry is obtained by
the leading order approximation to this model \cite{gh}. The corrections to the Newtonian gravity have also been studied using a six
dimensional warped brane world model \cite{si}. In this  model, apart from corrected Newtonian potential,
a non-linear gravitational wave solution has also been studied.
The   lowest order quantum corrections to a real scalar field in the Randall-Sundrum models have also been analyzed. It has been demonstrated that these type of corrections can have implications for
the  four dimensional  gravity  \cite{si2}. Thus, there are various kinds of modifications of the Newtonian law. It would be interesting to analyze the measurement of these modified potentials, by analyzing the modified dynamics from these corrected Newtonian potentials.

\section{Acknowledgements}

We gratefully acknowledge ingenious suggestions from Murray Neuman at University of British Columbia, Okanagan. We also acknowledge valuable insights from Anil Kumar at Indian Institute of Science Education and Research, Bhopal. We would like to extend our gratitude to Lan Fu at Australian National University for her discernment and enlightening inputs. We would also like to acknowledge an anonymous referee for their constructive feedback. 

\section{Appendix A: Derivation of parameters in the damping scenario}

\begin{figure}[h!]
\begin{center}
$%
\begin{array}{cccc}
\includegraphics[width=80 mm]{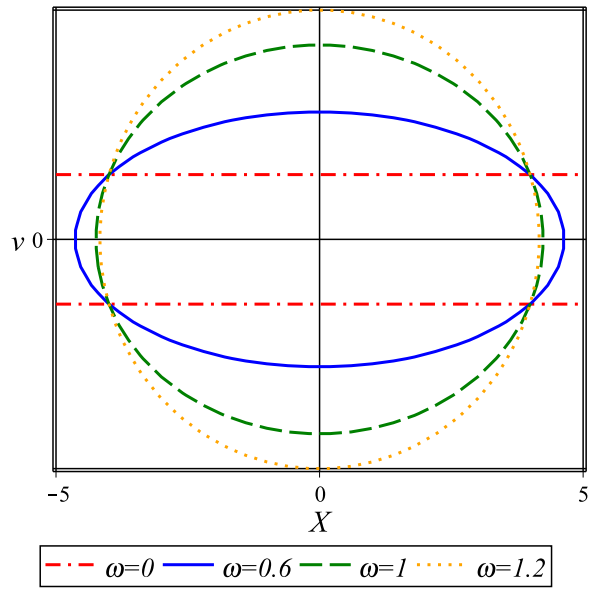}
\end{array}%
$%
\end{center}
\caption{Phase diagram for $k=c=1$.}
\label{fig2}
\end{figure}

Vanishing $t$ between $X=x+4k$ and $v$ gives phase diagram which is illustrated by the Fig. \ref{fig2}. In the case of $\omega=1$, the phase diagram is a circle with radius $\sqrt{16k^2+2c}$.

\begin{figure}[h!]
\begin{center}
$%
\begin{array}{cccc}
\includegraphics[width=65 mm]{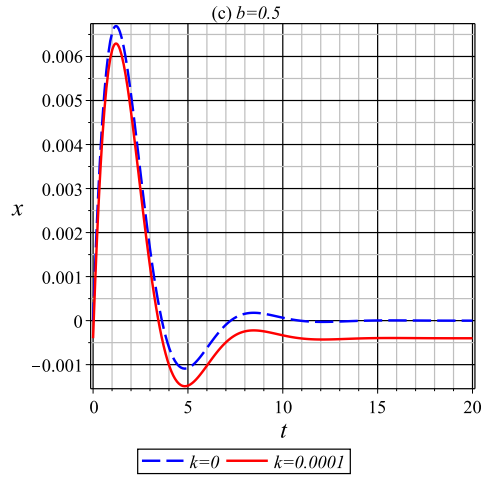}
\end{array}%
$%
\end{center}
\caption{Motion of particle versus time for $c=0.5$ and $\omega=1$.}
\label{fig3}
\end{figure}

Now  to measure these correction, we will  analyze the effects of drag force on  the system. So, we consider a drag force proportional to the velocity, for which the equation for this system can then be written as
\begin{equation}\label{11}
\frac{d^2x}{dt^2} = -\omega^2x-4\omega^2k-2b\frac{dx}{dt},
\end{equation}
where $b$ is the drag coefficient. Now we can write the solution to this equation as 
$
x = e^{-bt}(C_{1}e^{\sqrt{b^{2}-\omega^{2}}t}+C_{2}e^{-\sqrt{b^{2}-\omega^{2}}t}) - 4k.
$
If $\omega>b$ then the particle motion is under damping, and is described by 
$
x = ({\omega})^{-1}(\sqrt{16k^2\omega^2+2c})e^{-bt}\sin(\omega_{1}t) - 4k,
$
where $\theta_{0}=0$ is set as before, and $\omega_{1}=\sqrt{\omega^{2}-b^{2}}$. In Fig. \ref{fig3}, we can see that the motion of the particle is periodic. We can see difference in the corrected and the classical (Fig. \ref{fig3}) cases by shift in position. The same result is obtained for the particle acceleration.

\begin{figure}[h!]
\begin{center}
$%
\begin{array}{cccc}
\includegraphics[width=65 mm]{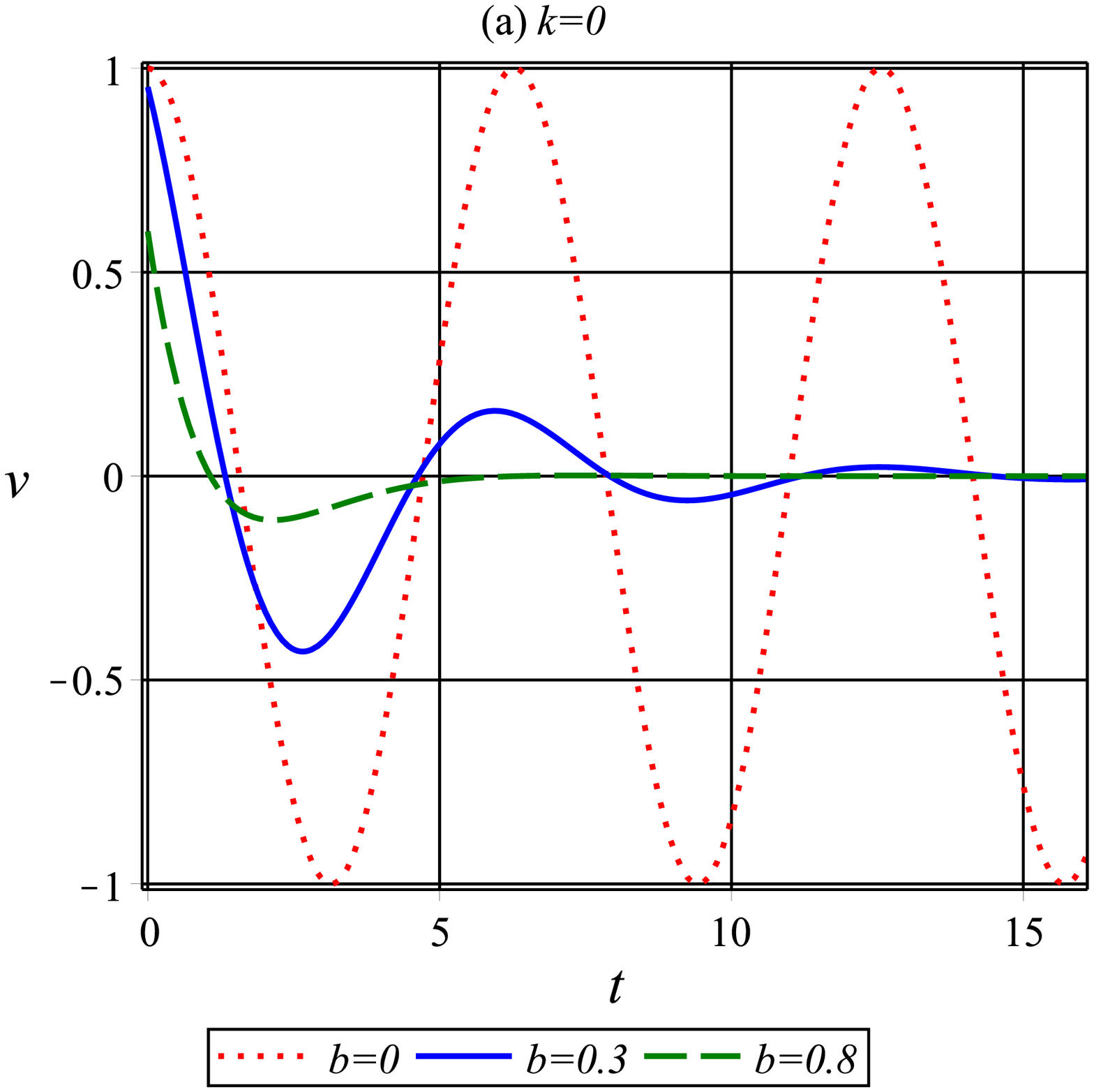}\includegraphics[width=65 mm]{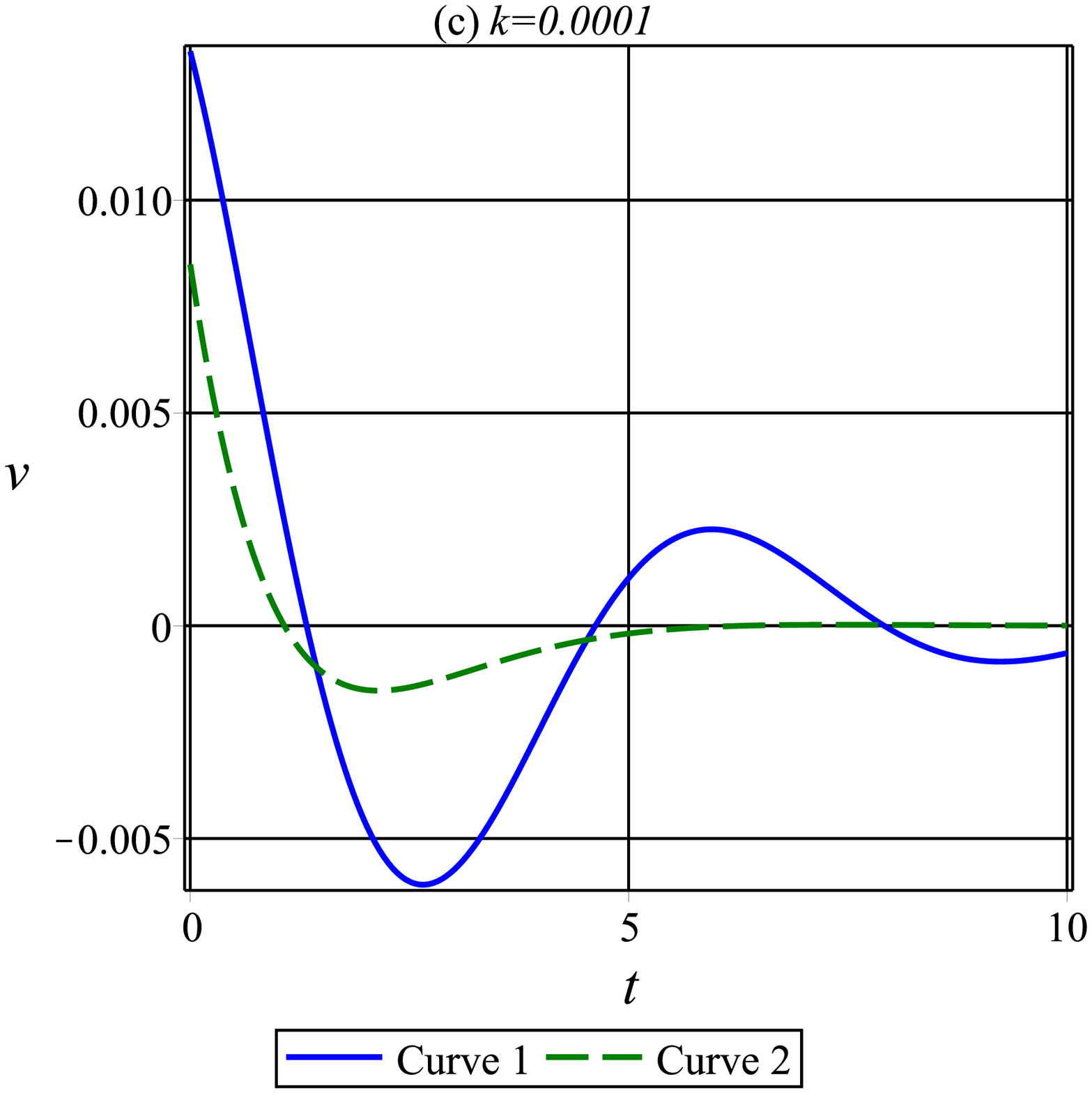}
\end{array}%
$%
\end{center}
\caption{Velocity of particle versus time for $c=0.5$ and $\omega=1$.}
\label{fig3-4}
\end{figure}

Now the velocity for the particle moving under modified Newtonian potential is given by 
$
v = ({\omega})^{-1}{(\sqrt{16k^2\omega^2+2c})e^{-bt}}\left[\omega_{1}\cos(\omega_{1}t)-b\sin(\omega_{1}t)\right].
$
Velocity in terms of time is plotted in Fig. \ref{fig3-4} which shows periodic features.
The phase equation of this case in polar coordinates is obtained as
$
r= ({\omega_{1}})^{-1} {\omega_{1}}{\omega}(\sqrt{16k^2\omega^2+2c})e^{-\frac{b}\theta},
$
with $\theta=\omega_{1}t$ and $r=\sqrt{(bX+v)^2+\omega_{1}^{2}X^{2}}$. It may be noted that the motion is represented by a  logarithmic spiral.

\end{document}